\documentclass[10pt,aps,prb,twocolumn,noshowpacs,superscriptaddress]{revtex4}
\usepackage{mathrsfs, bm,bbm,color,amsmath,amssymb, stmaryrd, amsfonts,latexsym,graphicx}
\usepackage{bbold,textcomp}
\usepackage{calrsfs}
\usepackage{color}
\usepackage{verbatim}
\usepackage{lmodern}
\usepackage[colorlinks=true]{hyperref}

\newcommand{\fred}{\color{black}}

\newcommand{\bs}[1]{\pmb{#1}}

\newcommand{\ud}{\,\mathrm{d}}
\newcommand{\Tr}{\mathrm{Tr}\,}
\newcommand{\tr}{\mathrm{tr}\,}
\newcommand{\ep}{\varepsilon}
\renewcommand{\k}{{\bm k}}
\renewcommand{\a}{{\bm a}}
\renewcommand{\r}{{\bm r}}

\newcommand{\h}{{\bm h}}
\newcommand{\n}{{\bm n}}
\newcommand{\dos}{{density of states}}
\newcommand{\id}{\mathbb{1}}
\newcommand{\be}{\begin{equation}}
\newcommand{\ee}{\end{equation}}
\newcommand{\ba}{\begin{array}}
\newcommand{\ea}{\end{array}}

\newcommand{\inter}{\mathrm{inter}}

\newcommand{\FV}{Fermi sea }
\newcommand{\FS}{Fermi surface }

\newcommand{\AND}{and }

\newcommand{\plateau}{\emph{plateau}}

\newcommand{\chiorb}{\chi_\text{orb}}
\newcommand{\chiLP}{\chi_\text{LP}}
\newcommand{\chig}{\chi_g}
\newcommand{\chigtilde}{\tilde{\chi}_g}
\newcommand{\chiOm}{\chi_\Omega}
\newcommand{\chiinter}{\chi_\text{inter}}
\newcommand{\chiL}{\chi_\text{L}}
\newcommand{\chigap}{\chi_\text{gap}}
\newcommand{\chiPauli}{\chi_\text{spin}}

\newcommand{\Vg}{Z_g}
\newcommand{\Vgtilde}{\tilde{Z}_g}
\newcommand{\VOm}{{\cal M}}

\newcommand{\dx}[1]{\,\partial_{x}#1}
\newcommand{\dy}[1]{\,\partial_{y}#1}
\newcommand{\dxx}[1]{\,\partial_{x}^2#1}
\newcommand{\dxy}[1]{\,\partial_{xy}^2#1}
\newcommand{\dyy}[1]{\,\partial_{y}^2#1}
\newcommand{\di}[1]{\,\partial_{i}#1}
\renewcommand{\dj}[1]{\,\partial_{j}#1}

\newcommand{\moy}[1]{\left\langle #1 \right\rangle_\text{BZ}}
\renewcommand{\emph}[1]{\textit{#1}}

\begin{document}

\title{Geometric orbital susceptibility: quantum metric without Berry curvature}
\author{Fr\' ed\' eric \surname{Pi\' echon}}
\email{piechon@lps.u-psud.fr}
\affiliation{Laboratoire de Physique des Solides, CNRS, Univ. Paris-Sud, Universit\'e Paris-Saclay, 91405 Orsay Cedex, France}
\author{Arnaud \surname{Raoux}}
\affiliation{Laboratoire de Physique des Solides, CNRS, Univ. Paris-Sud, Universit\'e Paris-Saclay, 91405 Orsay Cedex, France}
\affiliation{D\' epartement de Physique, \' Ecole Normale Sup\'erieure, PSL  Research  University, 24 rue Lhomond, 75005 Paris, France}
\author{Jean-No\" el \surname{Fuchs}}
\affiliation{Laboratoire de Physique des Solides, CNRS, Univ. Paris-Sud, Universit\'e Paris-Saclay, 91405 Orsay Cedex, France}
\affiliation{Laboratoire de Physique Th\' eorique de la Mati\` ere Condens\' ee, CNRS, Sorbonne Universit\'es, Universit\'e Pierre et Marie Curie, 4,
place Jussieu, 75252 Paris Cedex 05, France}
\author{Gilles \surname{Montambaux}}
\affiliation{Laboratoire de Physique des Solides, CNRS, Univ. Paris-Sud, Universit\'e Paris-Saclay, 91405 Orsay Cedex, France}

\date{\today}

\begin{abstract}
The orbital magnetic susceptibility of an electron gas in a periodic potential depends not only on the zero field energy spectrum 
but also on the geometric structure of cell-periodic Bloch states which encodes interband effects. 
In addition to the Berry curvature, we explicitly relate the orbital susceptibility of two-band models to  
a {\it quantum metric tensor} defining a distance in Hilbert space. Within a simple tight-binding model allowing 
for a tunable Bloch geometry, we show that interband effects are essential {\it even in the absence 
of Berry curvature}. We also show that for a flat band model, the quantum metric gives rise to a very strong orbital paramagnetism.
\end{abstract}


\maketitle

\section{Introduction}
The orbital susceptibility\cite{AM} 
measures the response of a time reversal invariant electronic system to an external magnetic field $B$ and is defined as the second derivative of the grand potential. 
Although being a thermodynamic quantity obtained in a perturbative limit ($B \rightarrow 0$), its evaluation is not simple, 
since it has been known for a long time that it depends not only on the zero-field band energy spectrum,\cite{Landau30,Peierls33} 
but also on the wavefunctions which encode interband effects.\cite{Adams52,Hebborn59,Roth61,Blount,Wannier64,Misra69,Fukuyama70,Fukuyama71}

One well-known quantity which describes interband effects is the Berry curvature.\cite{Xiao10} 
For example, it enters --\,together with the orbital magnetic moment\,-- in the expression of the magnetization, 
the first derivative of the grand potential with respect to the magnetic field.\cite{Thonhauser05,Thonhauser11,DiXiao05}  
It also appears in the expression of the susceptibility, but other geometrical quantities are expected since the susceptibility
is  a second derivative with respect to the magnetic field.\cite{Blount,Gao15} Here  we explicitly relate the susceptibility 
to the {\it quantum geometric tensor} introduced by M.V. Berry, whose imaginary part is the Berry curvature and whose 
real part is the so-called {\it quantum metric tensor}.\cite{Berry89,Vallee} 
The main goal of this paper is to show the central role played by this metric tensor on the structure of the orbital susceptibility.

Until recently the metric tensor was considered as a theoretical object useful to characterize 
 the localization properties Wannier functions in band insulators \cite{Marzari-Vanderbilt,RestaEPJB11}. However few recent works have suggested 
different physical properties such as current noise and superfluid weight that depend on the quantum metric in an essential way \cite{Neupert13,PeanoTorma15,Inamoglu15}. 
Moreover, in artificial crystals made of cold atoms, a full reciprocal space map of the quantum metric should be accessible via St\"uckelberg interferometry \cite{Lim15}. 
 

In a recent paper, we derived a general formula for the orbital susceptibility within a tight-binding picture (restricted here to $d=2$ dimensions):\cite{RaouxPRB15}
\begin{equation}
\chiorb(\mu,T)=-\frac{\mu_0 e^2}{12\hbar^2}\frac{\Im m}{\pi S}\int_{-\infty}^{\infty} n_\mathrm{F}(E)\Tr \hat X \ud E \ ,
\label{eq:chi}
\end{equation}
where the operator $\hat X$ is written in terms of the zero-field Green function $\hat g$ and of the derivatives $\dx$ and $\dy$ of the Bloch Hamiltonian 
$\hat h(\k)$ with respect to the components $k_x$ and $k_y$ of the wavevector:
\begin{equation}
\hat X = \hat g \dxx{\hat h}\,\hat g \dyy{\hat h}-\hat g \dxy{\hat h}\,\hat g \dxy{\hat h}+2 ([\hat g \dx{\hat h}, \hat g \dy{\hat h}] )^2 \ .
\end{equation}
Moreover, the orbital susceptibility was shown to satisfy a general sumrule over the full bandwidth:\cite{Gomez-Santos11,RaouxPRL14,Stauber15}
\begin{equation}
 \label{sumrule}
 \int \chiorb(\mu,T)\ud\mu=0\ .
\end{equation}
Although Eq.~\eqref{eq:chi} is complete (checked against numerical calculations for various models\cite{RaouxPRB15}), it hides many subtle effects 
that we wish to discuss here in the simplest context of two-band models. 

{\fred{
The outline of this paper is as follows: section II presents the quantum geometric properties that characterize the cell-periodic Bloch states $|u_\alpha(\k)\rangle$ 
of a given energy band $\epsilon_{\alpha}(\k)$. In particular, in addition to the well-known Berry curvature tensor $\Omega_{\alpha ij}$
we also introduce the quantum metric tensor $g_{\alpha ij}$ as a measure of the distance between Bloch states. 
Section III presents a general formula for the orbital susceptibility $\chiorb$ of two-band models. It is shown that, in contrast to
the intraband Landau-Peierls contributions $\chiLP$ which depends only on the energy band spectrum, the interband contribution $\chiinter$ crucially depends 
on the quantum geometry of Bloch states. More precisely $\chiinter$ may be decomposed in three contributions $\chiinter =  \chiOm + \chig +\chigtilde$ where
$\chiOm$ depends only on the Berry curvature whereas $\chig$ and $\chigtilde$ depend only on the quantum metric. 
Details of the derivation are given in Appendix A.
In section IV  we present explicit calculations of the different orbital susceptibility contributions for particular models that were designed
in order to highlight the physics hidden in the three interband geometric contributions.
Section V provides a more heuristic derivation and also suggests a possible qualitative interpretation for each of the three interband geometric contributions.
In particular $\chiOm$ is interpreted as a measure of the $\k$-space fluctuations of the spontaneous orbital magnetization whereas $\chig$ and $\chigtilde$ are interpreted 
as field induced effects resulting from the field induced horizontal and vertical positional shifts \onlinecite{Gao14}.
Appendix B provides more details on these positional shifts. 
In section VI we explain how our formulation compares with previous works; more precisely we discuss Blount's formula \cite{Blount} 
and also the more recent susceptibility formula obtained by Gao et al \cite{Gao15}.
The paper ends with a conclusion and perpspectives.
}}

\section{Geometry: quantum metric and Berry curvature} 

In order to describe the evolution of a cell-periodic Bloch state $|u_\alpha(\k)\rangle$ 
under the variation of a vector parameter~$\k$, M.V. Berry introduced the {\it quantum geometric tensor}~$T_{\alpha}$:
\begin{equation}
T_{\alpha ij}(\k)= \langle \partial_i u_\alpha | 1 -{\cal P_\alpha}  |\partial_j u_\alpha \rangle\ , 
\end{equation}
where ${\cal P_\alpha}(\k) = | u_\alpha  \rangle \langle u_\alpha |$ is the projector on the band $\alpha$ of energy $\varepsilon_{\alpha}(\k)$.
The imaginary (antisymmetric) part of $T_{\alpha ij}$ is nothing but the Berry curvature tensor: 
$\Omega_{\alpha ij}(\k)=- 2 \,  \mbox{Im}\,    T_{\alpha ij}$. 
The real (symmetric) part, named the {\it quantum metric tensor} $g_{\alpha ij}$,  characterizes a distance in Hilbert space, defined as:\cite{Vallee}
\be
ds_\alpha^2 \equiv 1 - |\langle u_{\alpha}(\k) |u_{\alpha}(\k+d\k)\rangle |^2  \ . 
\end{equation}
Expanding the $\k$ dependence of the wave functions to second order, the   tensor   $g_{\alpha ij}$ is defined as
\begin{equation}
ds_\alpha^2 = g_{\alpha ij} \ud k^i \ud k^j  \quad \mbox{with} \quad   g_{\alpha ij}(\k)= \mbox{Re} \ T_{\alpha ij} \ .  
\end{equation}
{\fred{The curvature and quantum metric tensors have the $\k$-space periodicity of the {\em reciprocal lattice} even if the Bloch states $|u_\alpha\rangle$ do not have it. 
Moreover, they stay invariant upon a {\em Berry gauge} tranformation $|u_\alpha \rangle  \rightarrow e^{i\varphi_\alpha(\k)} |u_\alpha \rangle$.
Systems with time reversal symmetry verify $\Omega_{\alpha ij}(-\k)=-\Omega_{\alpha ij}(\k)$ and $g_{\alpha ij}(-\k)=g_{\alpha ij}(\k)$.
Centro-symmetric systems verify $\Omega_{\alpha ij}(-\k)= \Omega_{\alpha ij}(\k)$.}}
In the following, to simplify further notations, the $\k$ dependence of quantities will be explicitly written only in their definitions.

We now restrict to two-band models. The $\bm k$-space Hamiltonian matrix can be written as 
\begin{align}
\label{hamiltonian}
\hat h(\k)&=\varepsilon_0(\k)\id+\bm h(\k)\cdot \bm\sigma=\varepsilon_0(\k)\id+\varepsilon(\k) \bm n(\k)\cdot \bm\sigma
\end{align}
where $\bm\sigma$ is the vector of Pauli matrices, and $\bm n(\k)$ a 3-dimensional unit vector depending on the $d$-dimensional vector $\bm k$. 
The Hamiltonian matrix has two eigenvalues $\varepsilon_{\alpha}(\k)=\varepsilon_0(\k)+\alpha \varepsilon(\k)$
with corresponding projectors ${\cal P_\alpha}(\k)=\frac{1}{2}(\id+\alpha\bm n\cdot \bm\sigma)$
where $\alpha=\pm$. In that situation, the Berry curvature and metric tensors components 
verify $\Omega_{\alpha ij} \equiv \alpha \Omega_{ij}$ and $g_{\alpha ij}\equiv g_{ij}$ with
\begin{equation}
\Omega_{ij}(\k) = \frac12 (\di{\n} \times \dj{\n}) \cdot \n\ ,\quad  g_{ij}(\k)= \frac14  \di\n \cdot \dj{\n}
\label{metric-2bandes} 
\end{equation}
{\fred{where each component of the curvature tensor verifies the identity 
\be
\Omega_{ij}^2  = 4 (g_{ii}g_{jj}-g_{ij}^2).
\label{omega-g}
\ee
In other words, for each vector $\k$, the quantum metric determines the modulus of the Berry curvature but not its $\k$ dependent sign.

\medskip
In this work we consider more specifically the case $d=2$.
In that situation the Berry curvature tensor has a single non-vanishing component $\Omega=\Omega_{xy}$ 
and the quantum metric tensor $g$ is a $2\times2$ symmetric matrix with the three elements $g_{xx},g_{yy},g_{xy}$.
In addition to the covariant metric tensor $g_{ij}(\k)$, it is convenient to further introduce a contravariant metric tensor $g^{ij}$ through
the identity $g^{ik}g_{kj}={\rm det}(g) \delta^i_j$ 
such that 
\be
(g^{xx},g^{yy},g^{xy})\equiv(g_{yy},g_{xx},-g_{xy}).
\label{contravariant}
\ee
The identity (\ref{omega-g}) can then be rewritten as
\be
\Omega^2  = 4 \,  \mbox{det} \, g= 4 g_{ij} g^{ij}.
\ee
The quantum geometric tensor of each band can then be written as
\be
T_{\alpha}(\k)=\frac{\Tr g}{2}(\id+\bm \tau_\alpha\cdot \bm \sigma)
\ee
with the unit vector 
\be
\bm \tau_\alpha(\k)\equiv \frac{1}{\Tr g}(2g_{xy},\alpha \Omega,g_{xx}-g_{yy})
\ee
For systems with time reversal symmetry the vector $\tau_\alpha(\k)$ has the same symmetry properties
as ${\bm n}(\k)$. 
Note that in artificial crystals made of cold atoms, 
a full reciprocal space map of the quantum metric tensor $T_{\alpha}(\k)$ should be accessible via St\"uckelberg interferometry \cite{Lim15}. 

}}

\section{Orbital susceptibility for two-band models}

We now analyse the different contributions to the orbital susceptibility{\fred{; the details of their explicit derivation are given in appendix A }}.
Quite generally, the orbital susceptibility can be decomposed in
\be
\chiorb = \chiLP + \chiinter
\ee
where the first term called Landau-Peierls (LP) 
only involves the zero field band spectrum 
whereas the second term, here referred to as interband, encodes all wavefunctions geometric effects.

\subsection{Landau-Peierls contribution}
The Landau-Peierls (LP) contribution $\chiLP$ writes 
(in units of $\frac{\mu_0 e^2}{\hbar^2}$):\cite{Peierls33}
\begin{equation}
\chiLP(\mu,T)=\moy{
\frac{n'_\alpha}{12} (\dxx\varepsilon_{\alpha} \dyy\varepsilon_{\alpha}- \dxy\varepsilon_{\alpha}    \dxy\varepsilon_{\alpha})
}
\label{chiLP} 
\end{equation}
with the shorthand notations used throughout the paper
\begin{equation}
n_\alpha \equiv n_F( \varepsilon_{\alpha}(\k)) \ , \   \moy{\,\bullet} \equiv
\sum_{\alpha=\pm}   \int \,\bullet \,  \frac{\ud^2 k}{4 \pi^2} \ 
\label{shortnotations}
\end{equation}
and where $n_F(\ep)$ is the Fermi factor. 
This LP contribution only involves the energy spectrum and its Hessian near the Fermi level.
At parabolic band edges, it reduces to Landau diamagnetism with the effective band mass.
By contrast, in the vicinity of a Van Hove singularity it is strongly paramagnetic because 
the spectrum exhibits a saddle point and therefore masses of opposite sign.\cite{Vignale91}
In the multiband case, $\chiLP$ verifies the sumrule~\eqref{sumrule} for each band separately.
Being a Fermi level property (see the $n'_\alpha$ factor in Eq.~\eqref{chiLP}), {\em the LP contribution vanishes in a gap}.

\subsection{Interband geometric contributions}
We now come to the  structure of the interband contribution $\chiinter$. 
{\fred{It may itself be decomposed in three contributions as
\begin{equation}
 \chiinter =  \chiOm + \chig +\chigtilde \,
 \label{eq:chi_deuxbandes_termes}
\end{equation}
which explicitly depend on either the Berry curvature or the metric tensor, and which separately obey the sumrule~\eqref{sumrule}:
\be
\int \rm{d}\mu \chiOm(\mu)=\int \rm{d}\mu \chig(\mu)=\int \rm{d}\mu \chigtilde(\mu)=0.
\ee
}}

The first contribution $\chiOm$ is written in terms of the Berry curvature $\Omega$:
\begin{equation}
\chiOm= \moy{\left(-n_\alpha'+\alpha \frac{n_\alpha}{\ep}\right) \VOm^2}\ , \quad \VOm=\ep \Omega\ .
\label{chiomega}
\end{equation}
%
The term proportinal to $n'_\alpha$  can be understood as the Pauli paramagnetic contribution of the orbital magnetic moment 
${\cal M}(\k)=\varepsilon \Omega$.\cite{Blount,Gao15,Koshino10} 
Being a \FS term, it vanishes in a band gap. The term proportinal to $n_\alpha$ is always diamagnetic. 
Moreover, being a Fermi sea term it gives rise to a \plateau{} in a band gap. 
Due to the absence of Berry curvature, $\chiOm$ vanishes in centro-symmetric systems.

The contribution $\chig$ may be seen as more fundamental since it
is related to the metric tensor, which never vanishes for coupled bands.
It may be compactly written as a pure Fermi sea term:
\begin{equation}
\chig = \moy{\left(-\alpha \frac{n_\alpha}{\varepsilon} \right) \Vg} \quad, \quad \Vg=\frac{1}{2}\dj\left(\ep^2\di{g^{ij}}\right)
\label{chig}
\end{equation}
{\fred{where $\Vg$ explicitly involves the contravariant metric tensor $g^{ij}$ that was defined in (\ref{contravariant}).}}
This quantity $\Vg(\k)$ changes sign in the BZ and verifies $\int \ud^2 k \Vg(\k)=0$.
As a consequence, $\chig(\mu)$ may exhibit a diamagnetic or paramagnetic \emph{plateau} in a band gap.

{\fred{The third contribution $\chigtilde$ only appears
in the absence of particle-hole symmetry. }}
It may also be written as pure Fermi sea contribution that depends on the metric tensor:
\begin{align}
&\chigtilde = \moy{\left(-\alpha \frac{n_\alpha}{\varepsilon} \right) \Vgtilde}\\
&\Vgtilde= g^{ij} \di\ep_{0}  \dj\ep_{0}+\alpha\ep\di\left(g^{ij}\dj\ep_0\right).
\label{chiprimeg}
\end{align}
The first term of $\Vgtilde(\k)$ is always positive and thus leads to a paramagnetic plateau in a gap. 
The second part, changes its sign with band index $\alpha$ and its BZ average vanishes in a gap.

\section{Examples}

This section presents explicit calculations of the different contributions to the orbital susceptibility. We discuss particular models that were designed
in order to highlight the physics hidden in the three interband geometric contributions and also to illustrate their quantitative importance.
We first consider a lattice model with particle-hole symmetry (such that $\chigtilde =0$) for which the relative importance of the interband contributions $\chiOm$ and $\chig$
can be tuned by a continuous parameter. We show in particular the importance of the geometric tensor, {\em even in the absence of Berry curvature}.
In a second example we discuss the important differences between a lattice model and its low energy counterpart. This remark underlines the approximations done 
when the spectrum is linearized near the edges of a band gap.
The last example concerns a lattice model that is inversion symmetric (such that $\chiOm =0$) but extremely particle-hole 
assymetric since it exhibits a flat band. This flat band gives rise to a huge contribution  $\chigtilde$.

For comparison purpose, we also present calculations of the corresponding spin contribution
which, in absence of spin-orbit coupling, is limited to the Pauli susceptibility $\chiPauli$ which is simply proportional to the zero-field density of states.

\subsection{From square to honeycomb-like lattice}

\begin{figure*}[t]
\begin{center}
\includegraphics[width=17.5cm]{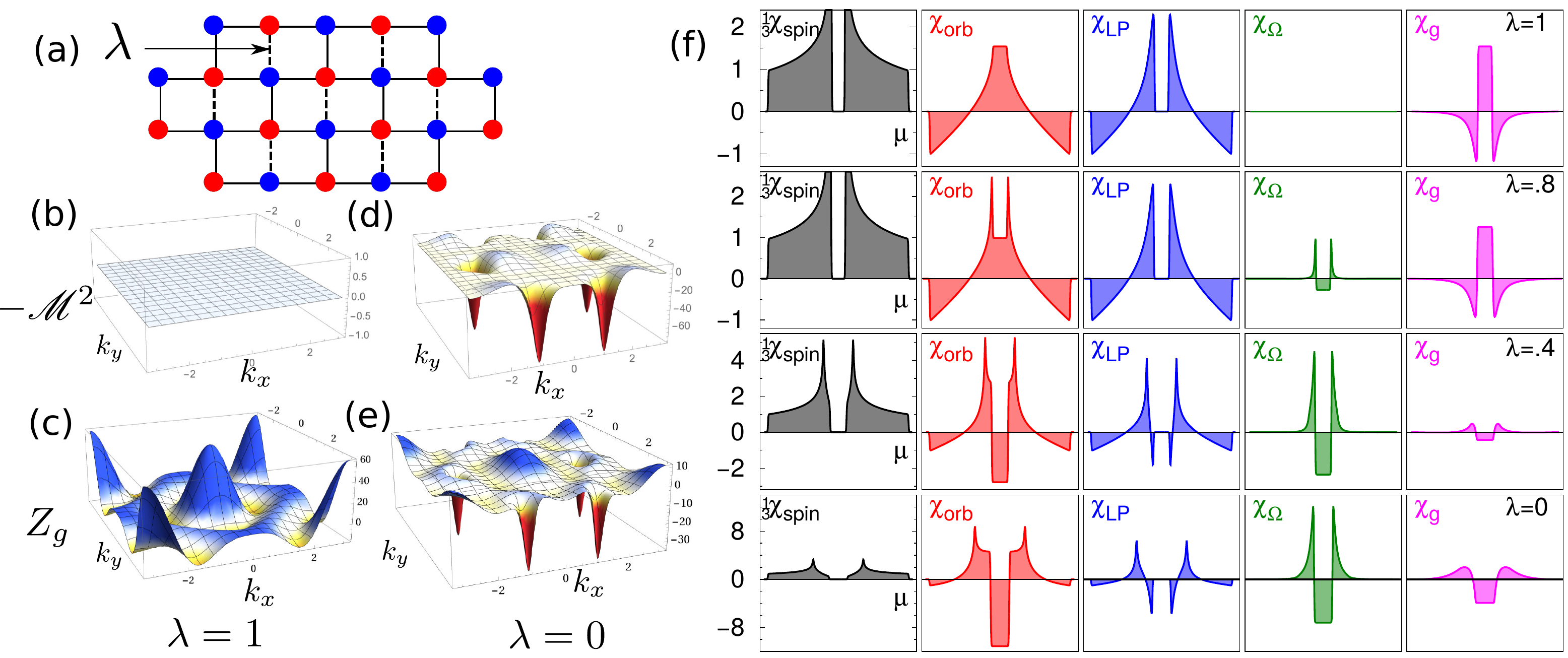}
\caption{(a) Tunable brickwall lattice with a staggered on-site potential $\Delta$: it interpolates between a square ($\lambda=1$) 
and a deformed honeycomb (brickwall) ($\lambda=0$). (b,d) Berry curvature term 
$-\VOm(\k)^2$ respectively for $\lambda=1$ and $0$. (c,e) Metric term $\Vg(\k)$ 
respectively for $\lambda=1$ and $0$. For clarity, the area of the plot covers twice the first Brillouin zone.
Blue and red colours denote respectively positive and negative values. (f) 
Evolution of the magnetic susceptibility versus chemical potential $\mu$ as a function of $\lambda$. 
The first column is the Pauli spin susceptibility 
which is proportional to the zero field \dos. It is normalized such that $\chiPauli=-3\chiorb$ 
at the band edges, as in the absence of a lattice. The four next columns concern the orbital response, respectively 
the susceptibility $\chiorb$ (red), $\chiLP$ (blue), $\chiOm$ (green) and $\chig$ (magenta). 
They are normalized to the Landau susceptibility $\chiL$ at the band edges, 
which itself depends on $\lambda$: $\chiL(\lambda)= \chiL(1) \sqrt{\frac{1 + 3\lambda}{3+\lambda}}$. 
The half bandwith equal to $\sqrt{\Delta^2+3 + \lambda}$, is also normalized to unity. We fix $\Delta/t=0.4$ and $T=0.001 t$.}
 \label{fig:square_lattice}
\end{center}
\end{figure*}


We consider a toy-model of electrons hopping on a tunable brickwall lattice with a staggered on-site potential $\Delta$. The  
tunable hopping parameter $\lambda$ interpolates between a square lattice ($\lambda=1$) 
where the Berry curvatuve is zero and a distorted honeycomb lattice ($\lambda=0$) 
where there is a finite Berry curvature concentrated in the vicinity of the Dirac points. 
The model is illustrated on Fig.~\ref{fig:square_lattice}(a).
Setting the nearest-neighbor coupling $t=1$ and the interatomic distance $a=1$, 
the corresponding Hamiltonian (cf. Eq.~\eqref{hamiltonian}) is given by the vector $\h(\k)= [2 \cos k_x + (1 + \lambda) \cos  k_y, (1 - \lambda) \sin k_y ,\Delta]$.

First, consider the case of the square lattice ($\lambda=1$).
The energy spectrum has two bands separated by a gap $2\Delta$. The density of states (proportional to the Pauli spin susceptibility $\chiPauli$ 
plotted in Fig.~\ref{fig:square_lattice}(f)) exhibits Van Hove singularities at gap edges.
The orbital susceptibility $\chiorb(\mu)=\chiLP+\chiOm+\chig$ is plotted on the top row of Fig.~\ref{fig:square_lattice}(f). 
In addition to the Landau diamagnetic behavior at the parabolic band edges, $\chiorb(\mu)$  exhibits a truncated logarithmic behavior at the gap edges, 
and a paramagnetic plateau in the gap. The Landau-Peierls contribution  $\chiLP$ properly exhibits Landau diamagnetism at the band edges, and also 
paramagnetic peaks reminiscent of the logarithmic divergence of the \dos \ at the gap edges. 
However it vanishes in the gap: therefore it cannot explain the paramagnetic plateau of $\chiorb$ in the gap.
For centro-symmetric systems there is no Berry curvature and therefore $\chiOm=0$. Thus, one naively expects the bands to be uncoupled. 
However, the quantum metric and $\Vg$ do not vanish, but give rise to a contribution $\chig$ that provides exactly the paramagnetic plateau in the gap. 
 As the gap goes to 0, this plateau diverges as $\log\Delta$. In addition, in order to respect the sumrule, $\chig$ presents also diamagnetic peaks near the gap edges. 
This striking example shows that {\it the interband coupling is not only encoded in the Berry curvature}, 
and that the metric contribution $\chig$ is essential to explain the structure of the total susceptibility.

We now consider the limit $\lambda=0$ defining the brickwall lattice. 
It has the same properties as the honeycomb lattice, except that the two Dirac points do not lie on symmetry lines of the BZ. 
It has been recently used in a cold-atom experiment to probe the existence, the motion and the merging of these Dirac points under proper variations of hopping parameters.
\cite{Tarruell:12,Lim:12} For this system, $\chiorb$ and its different contributions are presented on the bottom row of Fig.~\ref{fig:square_lattice}(f). 
In addition to the diamagnetic Landau regime at the band edges and the paramagnetic divergence at the Van Hove singularity, 
it exhibits a deep diamagnetic plateau in the gap, whose amplitude scales as $1/\Delta$, reminiscent of the diamagnetic $\delta$-peak found by McClure in the limit 
$\Delta \rightarrow 0$.\cite{McClure56,Koshino10} As it is well-known for the honeycomb lattice, there is a finite Berry curvature concentrated near the two Dirac points,
with opposite signs in the two valleys \cite{Fuchs10}, leading to $-\VOm^2$ plotted in Fig.~\ref{fig:square_lattice}(d).
In this situation, the quantity $\Vg(\k)$ in Fig.~\ref{fig:square_lattice}(e) shows also large negative peaks near the Dirac points, leading to a diamagnetic $\chig$ in the gap. 
Therefore the diamagnetic plateau is due both to the Berry curvature and the quantum metric contributions. 
Also remarkable is the paramagnetic \plateau{} near the edges of the gap.\cite{Gomez-Santos11,RaouxPRB15} This \plateau{} arises 
from the subtle compensation of a diverging diamagnetic peak of $\chiLP$, a diverging paramagnetic peak of $\chiOm$ and a smooth paramagnetic contribution $\chig$.

Figure ~\ref{fig:square_lattice}(f) 
presents the evolution of the susceptibility and its different contributions (which all obey separately the sumrule~\eqref{sumrule}) when varying $\lambda$ from 1 to 0. 
The contribution related to the Berry curvature, which is zero when inversion-symmetry is preserved, monotonously decreases in the gap and increases outside. 
The geometric contribution $\chig$ is more involved: inside the gap it evolves from a paramagnetic to a diamagnetic plateau 
whereas it has the reverse tendency outside the gap such as to respect the sumrule.

\subsection{Lattice model versus low energy model}

This section provides a quantitative comparison of the different 
susceptibility contributions between the brickwall lattice model at $\lambda=0$ and the corresponding 
linearized low energy effective model in the vicinity of the gapped Dirac points. It is interesting since it emphasizes the approximations 
which are made when linearizing the graphene
electronic spectrum (similar to the brick-wall lattice) in the vicinity of the gapped Dirac points.

The lattice model at $\lambda=0$ is given by $h(\k)=[2\cos k_x+\cos k_y,\sin k_y,\Delta]$. 
The gapped Dirac points are located at $(k_x,k_y)=(\xi \frac{2\pi}{3},0)$ with valley index $\xi=\pm1$.
The linearized model describing the vicinity of a Dirac point is given by $h_\xi(\k)=[\xi v_x k_x, v_y k_y,\Delta]$ with $(v_x,v_y)=(\sqrt{3},1)$.

For this linearized model it is then straightforward to obtain the equalities
\be
3{\cal H}={\cal M}^2=Z_g=\frac{\Delta^2 v_x^2 v_y^2}{4\varepsilon^4},
\ee
where ${\cal H}(\k)=\frac{1}{12}[\partial^2_{xx}\varepsilon\partial^2_{yy}\varepsilon-(\partial^2_{xy}\varepsilon)^2]$ and $\varepsilon(\k)=\sqrt{(v_x k_x)^2+(v_y k_y)^2+\Delta^2}$.
From these equalities, it is also immediate to deduce the explicit analytical form of the different susceptibility contributions in each valley. 
This is summarized in the following table ($\chi_0=\frac{1}{8\pi}\frac{v_x v_y}{\Delta}$):
\begin{center}
 \begin{tabular}{|c|c|c|}
   \hline
   & $|\mu|<\Delta$&$|\mu|>\Delta$\\
   \hline
$\chiLP$& $0$&$-\frac{1}{3}\chi_0\frac{\Delta^3}{\mu^3}$\\
$\chiOm$& $-\frac{1}{3}\chi_0$&$ +\frac{2}{3}\chi_0\frac{\Delta^3}{\mu^3}$\\
$\chig$& $-\frac{1}{3}\chi_0$&$-\frac{1}{3}\chi_0\frac{\Delta^3}{\mu^3}$\\
$\chiorb$ &$-\frac{2}{3}\chi_0$&$0$ \\
\hline
\end{tabular}
\end{center}
The expression for $\chiorb$ corresponds to a diamagnetic plateau in the gap and zero outside the gap. This coincides exactly with what has been derived directly from the Landau levels 
of a gapped Dirac spectrum \cite{Koshino10}. The present perturbative approach however provides the supplementary information that the apparent vanishing of $\chiorb$ oustside 
the gap results from the fortuituous compensation of the three contributions $\chiLP,\chiOm$ and $\chig$, 
since each of them is a power-law decreasing function outside the gap.

The quantitative comparison of the expressions for the low energy model with the exact lattice calculations are shown on Fig. 2.
One striking feature is that, for the contribution $\chiOm$, the lattice model and the low energy model calculations almost coincide.
In fact, the contribution $\chiOm$ of the linearized model verifies the sumrule.
By contrast, for the contribution $\chig$ there is a large quantitative difference for the value of the diamagnetic plateau between the lattice model and the low energy model. 
This difference explains quantitatively the value of the paramagnetic plateau (at gap edges) of $\chiorb$ for the lattice model. 
To summarize, we have shown that if $\chiOm$ essentially depends on the vicinity of the Dirac points, 
accounting quantitatively for $\chig$ requires to consider the whole Brillouin zone.

\begin{figure}
\begin{center}
\includegraphics[scale=0.75]{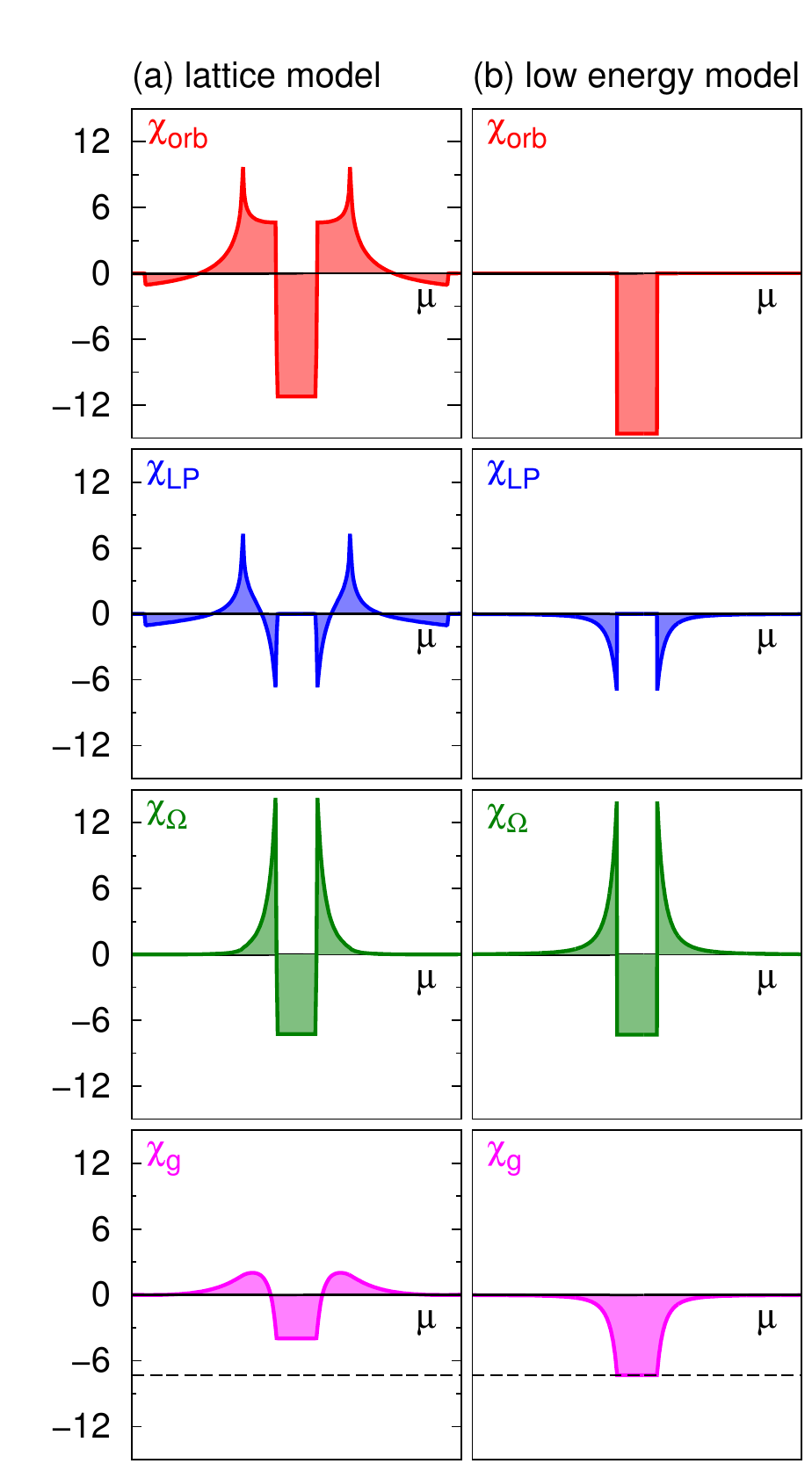}
\caption{(a) the susceptibility contributions $\chiorb$, $\chiLP$, $\chiOm$ and $\chig$ as a function of the chemical potential $\mu$ for the lattice model 
(this is identical to Fig 1. f with $\lambda=0$). 
(b) similar quantities but for the low energy (linearized) model.}
\end{center}
\end{figure}

\subsection{Mielke's checkerboard lattice model} 

%
%
%
\begin{figure}
\begin{center}
\includegraphics[scale=0.5]{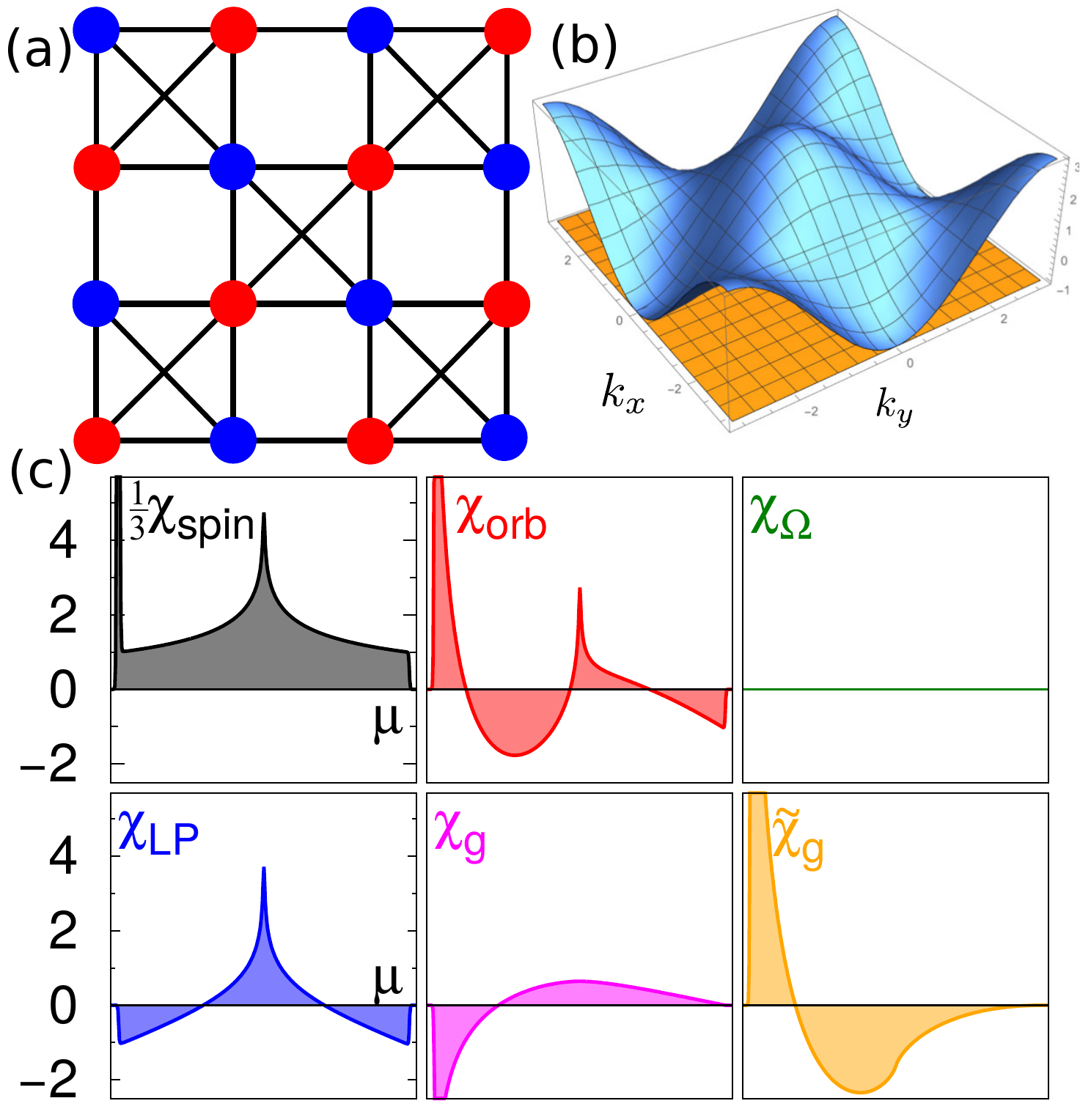}
 \caption{Top: (a) the checkerboard lattice (here there is no on-site potential) and (b) the corresponding energy spectrum. 
 Bottom: (c) Pauli spin susceptibility and orbital susceptibility $\chiorb$ with its different contributions as a function of the chemical potential. Units for $\chi$ and $\mu$ are similar to Fig. 
 \ref{fig:square_lattice}(f).}
 \label{fig:flats}
\end{center}
\end{figure}
%
As an illustration of a non particle-hole symmetric system, we consider a toy-model exhibiting a flat band. 
This is the Mielke checkerboard lattice shown on Fig.~\ref{fig:flats}(a), where all hopping integrals are identical ($t=\frac{1}{2}$). \cite{Mielke91,Aoki96}
The corresponding Hamiltonian is characterized by the vector $\h(\k)=(\cos k_x + \cos k_y,0, \sin k_x \sin k_y)$ 
and the energy  $\ep_{0}(\k)= \ep(\k) - 1$ (see Eq.~\eqref{hamiltonian}). The spectrum, shown in Fig.~\ref{fig:flats}(b), 
consists of a flat band touching the bottom of a dispersion relation which is that of the square lattice. 
One could expect naively the flat band to be inert in a magnetic field and the susceptibility to be simply given 
by the LP response of the square lattice (blue curve in Fig.~\ref{fig:flats}(c)). The exact result (red curve) is dramatically different, 
showing the importance of interband effects in this case.  
Because of inversion-symmetry, there is no Berry curvature and $\chiOm=0$ and therefore only the metric dependent terms $\chig$ and $\chigtilde$
can account for the strong interband effects. The most striking feature is the diverging paramagnetic peak of $\chiorb$ 
when approaching the energy of the flat band. Away from the flat band, this peak appears partially compensated by a wide diamagnetic shoulder. 
These two features come respectively from the first and second term in $\chigtilde$. In fact,
the contribution $\chig$ appears completely shadowed by that of $\chigtilde$. 
Both contributions show however that interband effects extend far away in energy from the flat band.
As a last remark, we note that Tasaki's two-band model on the square lattice \cite{Tasaki92,Aoki96},
which exhibits a flat band separated by a finite gap from the dispersive band,
gives rise to vanishing contributions $\chigtilde$ and $\chiOm$ and a finite contribution $\chig$ with a paramagnetic plateau in the gap.

{\fred{
\section{Heuristic derivation and interpretation of interband geometric contributions}

This section presents an heuristic derivation and an interpretation of the three interband susceptibility contributions $\chiOm,\chig,\chigtilde$.
To this end, it appears instructive to recall known results for the spontaneous orbital magnetization.

For generic multiband systems, the spontaneous orbital magnetization is given by  \cite{Xiao10}
\be
{M}(\mu,T)= 
\left\langle \left[n_\alpha {\mathcal {\bm M}}_\alpha
+ T\ln(1+e^{-(\varepsilon_\alpha-\mu)/T}) \Omega_\alpha\right]\right\rangle_{BZ}
\label{supp-mag}
\ee
where ${\mathcal {\bm M}}_\alpha(\k)$ is the orbital magnetic moment and $\Omega_\alpha(\k)$ is the Berry curvature of the band of energy $\varepsilon_\alpha(\k)$.
For time reversal invariant systems considered here, the spontaneous orbital magnetization vanishes because ${\mathcal {\bm M}}_\alpha(\k)=-{\mathcal {\bm M}}_\alpha(-\k)$ and
$\Omega_\alpha(\k)=-\Omega_\alpha(-\k)$. For a system that breaks time reversal symmetry, ${M}(\mu)$ 
can be nonzero however it has to verify the sum rule $\int d\mu \ M(\mu)=0$, valid for any multiband tight-binding model.
Although formula (\ref{supp-mag}) has been demonstrated in various ways,\cite{Xiao10,RaouxPRB15}
a useful heuristic derivation or reinterpretation consists of differentiating the grand canonical potential
\be
F(\mu,B,T)=-T\int d\epsilon \rho(\epsilon,B) \ln(1+e^{-(\epsilon-\mu)/T}),
\ee
to first order in magnetic field, with an effective magnetic field dependent density of states of the form \cite{Xiao10}
\be
\rho(\epsilon,B)=
\left\langle (1+B\Omega_\alpha)
\delta(\epsilon-\varepsilon_\alpha+{\mathcal {\bm M}}_\alpha B)\right\rangle_{\textrm{BZ}}.
\label{supp-dosmag}
\ee
The factor $(1+B\Omega_\alpha)$ is interpreted as a correction to the phase-space integration measure 
and $-{\mathcal {\bm M}}_\alpha B$ as a Zeeman-like correction of orbital origin to the band energy at first order in magnetic field.
To finish with the spontaneous orbital magnetization, we note that for two-band systems 
 ${\mathcal {\bm M}}_\alpha=\alpha \varepsilon \Omega_\alpha$ and $\Omega_\alpha=\alpha \Omega$, such that
 by introducing the two-dimensional Berry connection vector ${\bm a}_\alpha(\k)=i\langle u_\alpha|\nabla_\k|u_\alpha\rangle$ and using the identity
$\Omega_\alpha=[\nabla_\k \times {\bm a}_\alpha]_z$, the spontaneous magnetization Eq.(\ref{supp-mag}) of two-band systems can be recast in the form
\be
{M}(\mu,T)= 
\left\langle n_\alpha (\alpha \varepsilon [\nabla_\k \times {\bm a}_\alpha]_z+[{\bm v}_\alpha \times {\bm a}_\alpha]_z)\right\rangle_{\textrm{BZ}}.
\label{mag1}
\ee
with ${\bm v}_\alpha(\k)=\nabla_\k \varepsilon_\alpha$ the band velocity and
where the second term in Eq.(\ref{mag1}) follows from an integration by part of the second term in Eq.(\ref{supp-mag})

We now come to the heuristic derivation of the three interband susceptibility contributions. 
The idea consists of finding the modified effective density of states, valid to second order in magnetic field that permits to obtain 
the three contributions $\chiOm,\chig,\chigtilde$ from the second order derivative of the grand potential.
For two band systems we obtain the following effective density of states:
\be
\rho(\epsilon,B)= \moy{ (1+B\Omega_\alpha)
\delta(\epsilon-\varepsilon_\alpha+{\mathcal {\bm M}}_\alpha B-\frac{1}{2}{\mathcal {\bm M}}_\alpha \Omega_\alpha B^2 )},
\label{dosomega}
\ee
where now $\Omega_\alpha(B)$ and ${\mathcal {\bm M}}_\alpha(B)$ are field dependent quantities given by
\begin{align}
 \Omega_\alpha(B)=[\nabla_\k  \times {\bm a}_\alpha]_z+\frac{B}{2} \alpha [\nabla_\k  \times  \tilde{\bm  a}_g]_z,\\
 {\mathcal {\bm M}}_\alpha(B)=\alpha \varepsilon
\left(\Omega_\alpha(B) +\frac{B}{2}\frac{[\nabla_\k  \times \varepsilon^2 {\bm a}_g]_z}{\varepsilon^2} \right),
\end{align}
where $\a_\alpha$ is the zero-field Berry connection and 
${\bm a}_g,\tilde{\bm  a}_g$ are the first order field induced corrections \cite{Gao14} that verify (see appendix B for more details)
\begin{align}
\Omega_g(\k)=[\nabla_\k \times {\bm  a}_g ]_z=-\frac{1}{2} \partial_i \partial_j g^{ij},\\
\tilde \Omega_g(\k)=[\nabla_\k \times \tilde{\bm  a}_g ]_z=- \partial_i (\frac{g^{ij}\partial_j \ep_{0}}{\varepsilon}).
\end{align}
Within this picture, $\chiOm$ is given by
\be
\chiOm=\left\langle \left(-n_\alpha' {\mathcal {\bm M}}_\alpha^2 + {n_\alpha} {\mathcal {\bm M}}_\alpha \Omega_\alpha \right)   \right\rangle_{\textrm{BZ}}.
\label{supp-chiomega2}
\ee
where ${\mathcal {\bm M}}_\alpha=\alpha \varepsilon \Omega_\alpha$ and $\Omega_\alpha=\alpha \Omega$ are the zero field orbital magnetic moment and Berry curvature.
The contribution $\chiOm$ is thus a quadratic function of the quantities ${\mathcal {\bm M}}_\alpha,\Omega_\alpha$ 
that appear linearly in the orbital magnetization formula Eq.(\ref{supp-mag}).
This suggests that a natural interpretation for $\chiOm$ is
a measure of the $\k$-space fluctuations of the spontaneous orbital magnetization (the $\k$-space average of which vanishes). 
By contrast, $\chig$ and  $\chigtilde$ depend linearly on respectively ${\bm a}_g$ 
and $\tilde {\bm a}_g$: 
\be
\ba{l}
{\chig}
= \left\langle n_\alpha \ \alpha \varepsilon \left(\frac{[\nabla_\k  \times \varepsilon^2 {\bm a}_g]_z}{\varepsilon^2}\right) \right\rangle_{\textrm{BZ}},\\
\chigtilde=\left\langle n_\alpha (\alpha \varepsilon [\nabla_\k \times \tilde {\bm a}_g]_z+[{\bm v}_\alpha \times \tilde {\bm a}_g]_z)\right\rangle_{\textrm{BZ}},
\ea
\ee
The form of $\chigtilde$ being identical to the equality (\ref{mag1}), it strongly suggests to intepret the quantity $\chigtilde B$ as an induced magnetization.
By extension, we suggest that $\chig B$ may be interpreted as a field induced orbital magnetization resulting from a field induced orbital magnetic moment  
$\alpha \varepsilon  \frac{B}{2} \frac{[\nabla_\k  \times \varepsilon^2 {\bm a}_g]_z}{\varepsilon^2}$.
}}

\section{Comparison with other works}

Many different approaches were developped to calculate the orbital susceptibility in multiband systems (see discussion in \onlinecite{RaouxPRB15,Savoie12}). 
Compact but abstract orbital susceptibility formulas were obtained using Green's function techniques.
Other approaches usually result in an orbital susceptibility that is composed of several contributions. 
The main object of this section is to discuss the different decompositions that were obtained for the interband contribution $\chi_{\inter}(\mu)$.
Before discussing specific works, it is worth mentionning some generic features of the interband contribution $\chi_{\inter}(\mu)$.
Quite generally $\chi_{\inter}(\mu)$ may be formally written as the sum of individual band contribution $\chi_{\inter,\alpha}(\mu)$
where each $\chi_{\inter,\alpha}(\mu)$ is composed of both \FS and \FV contributions.
For each band, the \FS term depends on $n'_\alpha$ and thus vanishes outside the energy band. By contrast
the \FV term proportional to $n_\alpha$ gives rise to a finite susceptibility plateau for all $\mu$ above the top edge of the $\alpha^{\textrm{th}}$ band. As
a consequence the integrated contribution $\int d\mu \ \chi_{\inter,\alpha}(\mu)$ is infinite. In retrospect, this result implies that the decomposition of 
$\chi_{\inter}(\mu)$ into individual band contribution $\chi_{\inter,\alpha}(\mu)$ is quite meaningless.
A natural issue is thus how to decompose $\chi_{\inter}(\mu)$ into several contributions $\chi_{\lambda}(\mu)$ that are independently meaningful.
From that perspective and as explained in the previous sections, in the present work $\chi_{\inter}(\mu)$ is decomposed into three contributions
$\chiOm,\chig$ and $\chigtilde$, that each verifies the following two properties: (i) it vanishes outside the full band spectrum, (ii)
it verifies the tight-binding sumrule $\int d\mu \ \chi_{\Omega,{{g}},\tilde {{g}}}(\mu)=0$, despite the fact that they all contain a \FV part.

In the following, the discussion focuses on two works that provide an explicit 
decomposition of the interband susceptibility $\chi_{\inter}$ into several distinct contributions.
The first one is Blount's pioneer work \cite{Blount} in which some interband contributions explicitly imply
 the orbital magnetic moment and the Berry curvature (note that the Berry curvature concept and terminology did not exist at that time.). 
 The second one is the recent semiclassical wavepacket approach developped in \cite{Gao14,Gao15} 
where it was shown that some interband contributions explicitly imply the quantum metric.

\subsection{Blount's decomposition of $\chiinter$}

More than fifty years ago, Blount \cite{Blount} proposed a decomposition of $\chiinter$ into six distinct contributions.
More precisely by defining $\chiinter(\mu)=-\frac{\partial^2}{\partial {\bf B}^2}F_{\inter}(\mu,{\bf B})$, Blount writes the grand potential 
$F=F_{\textrm{LP}}+F_{\textrm{inter}}$ with
\be
F_{\textrm{inter}}=F_{\textrm{Pauli}}+F_{\Omega}+F_{\textrm{VV}}+F_{\textrm{at}}+F_{\textrm{pa}}+F_{7},
\ee
and where each contribution is given by
\begin{widetext}

\be
\ba{l}
F_{\textrm{Pauli}}=\left\langle n'_\alpha \ \frac{1}{2}({\bm B}\cdot {\bs {\cal { M}}}_\alpha)^2\right\rangle_{\textrm{BZ}},\\
F_{{\Omega}}=\left\langle -n_\alpha  \ \frac{3}{4} ({\bm B}\cdot {\bs {\cal { M}}}_\alpha) ({\bm B}\cdot{\bm \Omega}_\alpha)\right\rangle_{\textrm{BZ}},\\
F_{\textrm{VV}}= \left\langle n_\alpha \ \sum_{\beta \ne \alpha} 
\frac{ |\sum_{\beta'\ne \alpha}{\bs {\cal { A}}}_{\alpha \beta'}\cdot ({\bm V}_{\beta'\beta}+{\bm v}_\alpha \delta_{\beta \beta'})|^2}
{\varepsilon_\alpha -\varepsilon_\beta}\right\rangle_{\textrm{BZ}},\\
F_{\textrm{at}}=\left
\langle n_\alpha  \ \frac{1}{2}  \sum_{\beta \ne \alpha}\frac{{\bs {\cal { A}}}_{\alpha \beta}\cdot {\bs {\cal { A}}}_{\beta \alpha}}{m}
 \right\rangle_{\textrm{BZ}},\\
F_{\textrm{pa}}=\left
\langle -n_\alpha  \ \frac{1}{2}  \sum_{\beta \ne \alpha} {({\cal { A}}_i)}_{\alpha \beta} {({\cal { A}}_j)}_{\beta \alpha}
\partial_{ij} ^2 \varepsilon_\alpha \right\rangle_{\textrm{BZ}},\\
F_{\textrm{7}}=\left\langle n'_\alpha \ \frac{1}{2}{\bm v}_{\alpha} \cdot \left[\sum_{\beta,\beta' \ne \alpha}(
{\bs {\cal { A}}}_{\alpha \beta} ({\bm V}_{\beta'\beta}+{\bm v}_\alpha \delta_{\beta \beta'})\cdot {\bs {\cal { A}}}_{\beta' \alpha}+c.c)
\right] \right\rangle_{\textrm{BZ}},
\label{blount1}
\ea
\ee
\end{widetext}
with the notations
\be
\ba{l}
{\bm V}_{\alpha \beta}=\langle u_\alpha| \nabla_\k \hat h|u_\beta\rangle,\\
{\bm v}_\alpha={\bm V}_{\alpha \alpha}=\nabla_\k \varepsilon_\alpha,\\
{\bm A}_{\alpha \beta}=\langle u_\alpha|i \nabla_\k |u_\beta\rangle=-i\frac{{\bm V}_{\alpha \beta}}{\varepsilon_\alpha-\varepsilon_\beta},\\
{\bs {\cal { M}}}_{\alpha}=\frac{1}{2} \sum_{\beta \ne \alpha} {\bm A}_{\alpha \beta} \times {\bm V}_{\beta \alpha},\\
{\bm \Omega}_{\alpha}=i \sum_{\beta \ne \alpha} {\bm A}_{\alpha \beta} \times {\bm A}_{\beta \alpha},\\
{\bs {\cal A}}_{\alpha \beta}=\frac{1}{2} {\bm B} \times {\bm A}_{\alpha \beta}.
\ea
\label{blount2}
\ee
All these expressions were derived starting from the Schr\"odinger Hamiltonian $\hat H=\frac{{\bm p}^2}{2m} +V(\r)$ in a periodic lattice potential $V(\r)$ 
such that $m$ is the bare electron mass.
The effect of the crystal lattice is implicitly taken into account through the existence of an infinite number of bands 
with effective dispersions relation $\varepsilon_\alpha (\k)$ and associated cell-periodic Bloch states $|u_\alpha \rangle$.

In Eqs (\ref{blount1}), the contribution $F_{\textrm{Pauli}}$ is \FS like and represents the effective Pauli paramagnetism 
of the orbital magnetic moment  ${\bs {\cal { M}}}_{\alpha}$  of the  $\alpha^{\rm{th}}$ band. 
The contribution $F_{{\Omega}}$ involves the product of the orbital magnetic moment by 
a quantity ${\bm \Omega}_{\alpha}$ which appears to be the Berry curvature. 
$F_{{\Omega}}$ is \FV like and Blount argued that it is diamagnetic. 
The third contribution represents the Van-Vleck paramagnetism of occupied bands; it involves interband geometric effect through
the interband Berry connection ${\bm A}_{\alpha \beta}$ or interband velocity operator ${\bm V}_{\alpha \beta}$.
The fourth and fifth terms are argued to constitute generalization of the Langevin atomic diamagnetism for electrons in occupied bands. 
In particular the fourth term is clearly diamagnetic whereas 
the sign of the fifth may change according to the band dispersion.
The last term $F_{\textrm{7}}$ is \FS like but its meaning remains unclear. 

Due to the very different starting point, it may appear quite difficult to compare Blount's results with the tight-binding approach, involving a finite number of band,
considered in this work.
Nevertheless it appears instructive to arbitrarily substitute the peculiar form of the Berry connection properties of two-band models into Blount's formula 
(for a magnetic field ${\bm B}$ perpendicular to a 2D plane). In doing this, each Blount's susceptibility contribution rewrites:
\be
\ba{l}
\chi_{\textrm{Pauli}}=\left\langle -n'_\alpha \ \varepsilon^2 \Omega^2\right \rangle_{\textrm{BZ}},\\
\chi_{{\Omega}}=\left\langle  n_\alpha  \  \frac{3}{2} \alpha \varepsilon \Omega^2\right \rangle_{\textrm{BZ}},\\
\chi_{\textrm{VV}}= \left\langle -n_\alpha \  g^{ij} \partial_{i}\varepsilon_0\partial_{j}\varepsilon_0 \right\rangle_{\textrm{BZ}},\\
\chi_{\textrm{at}}=\left\langle n_\alpha \  \frac{1}{4 m} g^{ij}\delta_{ij} \right\rangle_{\textrm{BZ}},\\
\chi_{\textrm{pa}}=\left\langle -n_\alpha \ \frac{1}{4} g^{ij} \partial_{ij} ^2 \varepsilon_\alpha \right\rangle_{\textrm{BZ}},\\
\chi_{\textrm{7}}= \left\langle n'_\alpha \  g^{ij} \partial_{i}\varepsilon_\alpha\partial_{j}\varepsilon_0 \right\rangle_{\textrm{BZ}}=
\left\langle -n_\alpha \  \partial_i(g^{ij} \partial_{j}\varepsilon_0 )\right\rangle_{\textrm{BZ}},
\label{blount3}
\ea
\ee
where an integration by part permits to rewrite $\chi_{\textrm{7}}$ also as a \FV term. 

By summing the two Blount contributions that depends on the Berry curvature it appears that $\chi_{\textrm{Pauli}}+\chi_{{\Omega}}$ 
is similar but differs by a numerical factor from the corresponding tight-binding contribution $\chiOm$ (\ref{chiomega}).
More precisely Blount's \FV term $\chi_{{\Omega}}$ is a factor $3/2$ bigger than in Eq.(\ref{chiomega}); 
as a result  $\chi_{\textrm{Pauli}}+\chi_{_{\Omega}}$ cannot verify the sum rule.
By contrast, it appears that Blount's contributions $\chi_{\textrm{VV}}$ and $\chi_{\textrm{7}}$ correspond perfectly to the first and second part of the contribution $\chigtilde$ 
in Eqs.(\ref{chig},\ref{chiprimeg}); as a result $\chi_{\textrm{VV}}+\chi_{\textrm{7}}$ verifies the sum rule.
The last two terms $\chi_{\textrm{at}}$ and $\chi_{\textrm{pa}}$ are however very different from the last contribution $\chig$ obtained for two-band tight-binding models.

From the above analysis one may conclude that the three contributions $\chi_{{\Omega}}$, $\chi_{{at}}$,  and $\chi_{{pa}}$ of Blount
need to be modified in order to recover two-band tight-binding formula.
In fact, it appears that only $\chi_{{at}}$ needs to be revised (see below). \cite{Gao15}
To conclude with Blount's formula, we note that for centro-symmetric systems ($\chi_{\textrm{Pauli}}=\chi_{\Omega}=0$), 
contributions identical to 
$\chi_{\textrm{VV}},\chi_{\textrm{at}},\chi_{\textrm{pa}}$ and $\chi_{\textrm{7}}$ 
were recently derived \cite{ogata2015} starting from the Fukuyama \cite{Fukuyama70,Fukuyama71} compact Green's function formula.

\subsection{Gao {\em et al}  \cite{Gao15} decomposition of $\chiinter$}

Using a semiclassical wavepackets method, Gao et al \cite{Gao15} recently presented a decomposition of the interband contribution $F_{\inter}(\mu)$ into five terms:
\be
F_\inter=F_{\textrm{Pauli}}+F_{\textrm{geom}}+F_{\textrm{Polar}}+F_{\textrm{VV}}+F_{\textrm{Langevin}},
\label{gaoniu0}
\ee
which are given by
\begin{widetext}
\be
\ba{l}
F_{\textrm{Pauli}}=\left\langle -n'_\alpha \ \frac{1}{2}({\bm B}\cdot {\bs {\cal { M}}}_\alpha)^2\right\rangle_{\textrm{BZ}},\\
F_{\textrm{geom}}=\left\langle n_\alpha  \ \left[(\frac{3}{4} ({\bm B}\cdot {\bs {\cal { M}}}_\alpha) ({\bm B}\cdot{\bm \Omega}_\alpha)
+\frac{1}{8}\epsilon_{sik}\epsilon_{tjl}B_s B_t g_{\alpha ij} \partial_{kl} ^2 \varepsilon_\alpha)\right] \right\rangle_{\textrm{BZ}},\\
F_{\textrm{Polar}}=\left\langle n'_\alpha \ \frac{1}{4}{\bm v}_{\alpha} \cdot {\bm P}_\alpha \right\rangle_{\textrm{BZ}}
,\\
F_{\textrm{VV}}= \left\langle -n_\alpha \ \sum_{\beta \ne \alpha} \frac{ {\bm G}_{\alpha\beta}
{\bm G}_{\beta\alpha}}{\varepsilon_\alpha -\varepsilon_\beta}\right\rangle_{\textrm{BZ}},\\
F_{\textrm{Langevin}}= \left\langle -n_\alpha \ \frac{1}{8}\left[
\sum_{\beta,\beta' \ne \alpha} ({\bf B} \times {\bm A}_{\alpha \beta})_i (\Gamma_{ij})_{\beta \beta'} ({\bm B} \times {\bm A}_{\alpha \beta})_j
-\frac{1}{2}({\bm B} \times {\bf \nabla})_i ({\bm B} \times {\bf \nabla})_j(\Gamma_{ij})_{\alpha \alpha} \right]\right\rangle_{\textrm{BZ}},
\label{gaoniu1}
\ea
\ee
where $\epsilon_{ijk}$ is the totally antisymmetric tensor and with
\be
\ba{l}
g_{\alpha ij}=\frac{1}{2}\sum_{\beta \ne \alpha} \left[({\bm A}_{\alpha \beta})_i ({\bm A}_{\beta \alpha})_j +c.c\right],\\
{\bm P}_\alpha=\frac{1}{4}\sum_{\beta,\beta' \ne \alpha} \left[( {\bm B}\times  {\bm A}_{\alpha\beta})({\bm v}_\alpha
\delta_{\beta \beta'}+ {\bm V}_{\beta\beta'})\cdot ({\bm B}\times  {\bm A}_{\beta'\alpha})+c.c\right],\\
G_{\alpha \beta}=
-\frac{1}{2} {\bm B}\cdot \left[\sum_{\beta' \ne \beta } 
({\bm V}_{\alpha \beta'}+{\bm v}_\beta \delta_{\alpha \beta'})\times {\bm A}_{\beta'\beta}\right],\\
({\Gamma}_{ij})_{\alpha \beta}=\langle u_\alpha|\partial^2_{ij} \hat h |u_\beta\rangle,
\label{gaoniu2}
\ea
\ee
\end{widetext}
where the quantities $g_{\alpha ij}$ and ${\bm P}_\alpha$ correspond respectively to
the quantum metric tensor and the polarization associated to the $\alpha^{\rm{th}}$ band; 
where in fact the polarization is argued to constitute another characteristic geometric quantity \cite{Gao15}. 

Despite the very different derivation and some change of notation, as compared to Blount, it is apparent that there are strong similarities between 
Blount's formula Eqs.(\ref{blount1}) and Gao {\em et al's} expressions Eqs.(\ref{gaoniu1}).
More precisely, it is easily shown that $F_{\textrm{Pauli}},F_{\textrm{VV}}$ are identical and moreover $F_{\textrm{geom}}=F_{{\Omega}}+F_{\textrm{pa}}$ and 
 $F_{\textrm{Polar}}=\frac{1}{2}F_{\textrm{7}}$. Apart from the {\em wrong} factor $1/2$ of $F_{\textrm{Polar}}$ (which has been already pointed out in \cite{ogata2015}), 
the main apparent difference between Blount and Gao {\em et al} formulations resides in the contribution $F_{\textrm{at}}$ 
for the former which becomes $F_{\textrm{Langevin}}$ in the latter. 
In fact when considering the same starting Hamiltonian (e.g. $\hat H=\frac{{\bm p}^2}{2m} +V(\r)$ in a periodic lattice potential $V(\r)$) 
it appears that $({\Gamma}_{ij})_{\alpha \beta}=\frac{1}{m}\delta_{ij} \delta_{\alpha \beta}$, with $m$ the bare electron mass, 
such that in this situation $F_{\textrm{Langevin}}=F_{\textrm{at}}$.

More interestingly, when considering generic two-bands tight-binding models, as in the present work, 
it is possible to show that in such a situtation the Langevin term gives rise to a 
susceptibility contribution of the form:
\begin{widetext}
\be
\chi_{\textrm{Langevin}}= \left\langle -n_\alpha \ \left[ \frac{1}{4} g^{ij} \partial_{ij} ^2 \varepsilon_\alpha + \frac{1}{2}\alpha \varepsilon \Omega^2+
\frac{1}{2}\alpha \frac{\partial_i(\varepsilon^2 \partial_j g^{ij})}{\varepsilon}\right]\right\rangle_{\textrm{BZ}}.
\label{gaoniu3}
\ee
Using this expression it is immediate to establish the following identities (valid for two-band models)
\be
\ba{l}
\chiOm+\chig=\chi_{\textrm{Pauli}}+\chi_{\textrm{geom}}+\chi_{\textrm{Langevin}},\\
\chigtilde=\chi_{\textrm{VV}} +2  \chi_{\textrm{Polar}},
\ea
\ee
\end{widetext}
where in the first line there is a complete cancellation of the term $\langle n_\alpha  \frac{1}{4} g^{ij} \partial_{ij} ^2 \varepsilon_\alpha \rangle_{\textrm{BZ}}$
present in both $\chi_{\textrm{geom}}$ and $\chi_{\textrm{Langevin}}$ but with opposite sign.
This cancellation is important because it can be checked that this term generically gives rise to an unphysical finite susceptibility plateau that extends to $\mu\rightarrow \infty$
when each contribution $\chi_{\textrm{geom}}$ and $\chi_{\textrm{Langevin}}$ are considered separately. 
An indication of this spurious plateau for each contribution is already visible 
on the Fig 2b of Gao {\em et al}\cite{Gao15}, despite the reduced range of $\mu$.

\section{Conclusion and perspectives}

This study shows the physical richness of the orbital susceptibility even in the simplest case of two coupled bands.
We have stressed that the Berry curvature $\Omega$ is not sufficient to describe interband effects which are still prominent even when $\Omega=0$, 
as shown for the square lattice with staggered potential (broken sublattice symmetry). A simple model in which inversion-symmetry is progressively broken 
shows how the complexity of the orbital response is driven by the structure of a quantum geometric tensor, whose antisymmetric part is the Berry curvature
and the symmetric  part ({\em the quantum metric}) carries information on the distance between Bloch states in Hilbert space.
The complex structure of the susceptibility
can be summarized by its value in the gap  which shows explicitly
the contributions related to the Berry curvature $\Omega$ 
and the metric tensor $g_{ij}$:
\be
\chigap = \moy{\frac{1}{\ep}(
-\ep^2\Omega^2+ \frac{1}{2} \di(\ep^2 \dj g^{ij})+ g^{ij} \di\ep_{0} \dj\ep_{0}
)}\ .
\nonumber
\ee
The surprising strong influence of a flat band even in the simplest case of a two-band model motivates further the study
 of flat band physics in multiband systems. On a more general perspective, 
the orbital susceptibility may be an important tool for the investigation of topological 
transitions in multiband systems since it  provides unique informations on  the evolution of the geometric
properties of Bloch states across such  transitions. 



%
%

\bigskip

{Acknowledgments : This work was supported by the french programs ANR DIRACFORMAG
(ANR-14-CE32-0003) and by the LabEx PALM Investissement d'Avenir under the grant (ANR-10-LABX-0039-PALM).}


\appendix

\begin{widetext}

\section{Orbital susceptibility of two-band models}

\subsection{Explicit derivation}

This section presents the main calculation steps of the four contributions $\chi_{\textrm{LP}}$, $\chi_\Omega$, $\chi_g$ and $\tilde \chi_g$.
To start with, the general susceptibility formula eq.(1) (main text) is rewritten in the form 
\begin{equation}
\chi_{\textrm{orb}}(\mu,T)=-\frac{\mu_0 e^2}{12\hbar^2}\frac{\Im m}{\pi S}\int_{-\infty}^{\infty} \ud E n_\mathrm{F}(E) 
\int_\mathrm{B Z} \frac{\ud^2k}{4\pi^2} (U(\k,E)+V(\k,E)).
\end{equation}
with
\be
\ba{l}
U(\k,E)=\tr\left\{(\hat g\partial_{xx}\hat h\hat g\partial_{yy}\hat h-\hat g\partial_{xy}\hat h\hat g\partial_{xy}\hat h)_{\k}\right\},\\
V(\k,E)=2 \ \tr\left\{([\hat g\partial_{x}\hat h,\hat g \partial_{y}\hat h]^2)_\k \right\}
\ea
\ee
where $\mathrm{tr}\{\bullet\}$ is the partial trace operator on the band index $\alpha=\pm$. 
By using similar steps as described in Appendix D of ref.\onlinecite{RaouxPRB15}, one obtains
\be
\begin{array}{l}
U=\sum_{\alpha} g_\alpha^2 U^{(1)}_{\alpha}+  g_\alpha g_{-\alpha} U^{(2)},\\
V=\sum_{\alpha} g_\alpha^3 g_{-\alpha} V^{(1)}_{\alpha}+ g_\alpha^2 g_{-\alpha}^2 V^{(2)}
\end{array}
\label{UV}
\ee
with $g_\alpha(\k,E)=\frac{1}{E-\varepsilon_{\alpha}(\k)}$ and where

\be
 \begin{array}{l}
U^{(1)}_{\alpha}(\k)
=( \partial_{xx} \varepsilon_{\alpha}\partial_{yy}\varepsilon_{\alpha}  - \partial_{xy}\varepsilon_{\alpha}\partial_{xy}\varepsilon_{\alpha}) 
+4(\varepsilon^2 \Omega^2-\alpha \varepsilon g^{ij}\partial_{ij}\varepsilon_{\alpha}),\\
U^{(2)}(\k)
=-4(\varepsilon^2 \Omega^2+\frac{1}{2}\partial_{ij}(\varepsilon^2g^{ij})- \varepsilon g^{ij} \partial_{ij}\varepsilon ),\\
V^{(1)}_{\alpha}(\k)
=-16 \epsilon^2 g^{ij}\partial_{i}\varepsilon_{\alpha}\partial_{j}\varepsilon_{\alpha},\\
V^{(2)}(\k)
=-16\varepsilon^2  (2\varepsilon^2 \Omega^2- g^{ij}\partial_{i}\varepsilon_{\alpha}\partial_{j}\varepsilon_{-\alpha}).
\end{array}
\ee
At this point, using the identity $\partial_{i}(g_\alpha^n)=ng_\alpha^{n+1}\partial_{i}\varepsilon_{\alpha}$ and integration by part,
the following identity is established:
\be
\int_\mathrm{B Z} \frac{\ud^2k}{4\pi^2} \sum_{\alpha} g_\alpha^3 g_{-\alpha} V^{(1)}_{\alpha}=
\int_\mathrm{B Z} \frac{\ud^2k}{4\pi^2} \sum_{\alpha} 8[g_\alpha^2 g_{-\alpha}^2 (\epsilon^2g^{ij}\partial_{i}\varepsilon_{\alpha}\partial_{j}\varepsilon_{-\alpha})+
g_\alpha^2 g_{-\alpha} (\epsilon^2g^{ij}\partial_{ij}\varepsilon_{\alpha})-\frac{1}{2}g_\alpha g_{-\alpha}\partial_{ij}(\varepsilon^2g^{ij})].
\ee
From there, by using the equalities
\be
\begin{array}{l}
g_\alpha g_{-\alpha}=\frac{\alpha}{2\varepsilon}(g_\alpha-g_{-\alpha}),\\
g_\alpha ^2 g_{-\alpha}^2=\frac{1}{4\varepsilon^2}(g_\alpha^2+g_{-\alpha}^2-2g_\alpha g_{-\alpha}),\\
g_\alpha ^2 g_{-\alpha}=\frac{\alpha}{2\varepsilon}(g_\alpha^2-g_\alpha g_{-\alpha}),
\end{array}
\ee
it is possible to rewrite Eq.(\ref{UV}) as
\be
\begin{array}{l}
U=\sum_{\alpha} g_\alpha^2 U^{(1)}_{\alpha}+\alpha \frac{g_\alpha}{\varepsilon} U^{(2)},\\
V=\sum_{\alpha} g_\alpha^2 {\tilde V}^{(1)}_{\alpha}+\alpha \frac{g_\alpha}{\varepsilon}{\tilde V}^{(2)}
\end{array}
\ee
with

\be
\begin{array}{l}
{\tilde V}^{(1)}_{\alpha}(\k)
=4(-4\varepsilon^2 \Omega^2+3  g^{ij}\partial_{i}\varepsilon_{\alpha}\partial_{j}\varepsilon_{-\alpha}+\alpha \varepsilon g^{ij}\partial_{ij}\varepsilon_{\alpha}),\\
{\tilde V}^{(2)}(\k)
=-4(-4\varepsilon^2 \Omega^2+3 g^{ij}\partial_{i}\varepsilon_{\alpha}\partial_{j}\varepsilon_{-\alpha} +\varepsilon  g^{ij}\partial_{ij}\varepsilon+\partial_{ij}(\epsilon^2g^{ij})).
\end{array}
\ee
Summing $U$ and $V$ yields
\be
\frac{ U+V}{12}=\sum_{\alpha} g_\alpha^2
[\frac{1}{12}( \partial_{xx}\varepsilon_{\alpha}\partial_{yy}\varepsilon_{\alpha}  - \partial_{xy}\varepsilon_{\alpha}\partial_{xy}\varepsilon_{\alpha}) 
-\varepsilon^2 \Omega^2+g^{ij}\partial_{i}\varepsilon_{\alpha}\partial_{j}\varepsilon_{-\alpha}]
+\alpha \frac{g_\alpha}{\varepsilon}
[\varepsilon^2 \Omega^2-g^{ij}\partial_{i}\varepsilon_{\alpha}\partial_{j}\varepsilon_{-\alpha} -\frac{1}{2}\partial_{ij}(\varepsilon^2g^{ij}) ].
\label{UVb}
\ee
The final step consists of performing the explicit integral over variable $E$ by using the identity
\be
\Im m\int_{-\infty}^{+\infty}\frac{n_\mathrm{F}(E)}{(E-\varepsilon_\alpha)^k}\ud E=-\frac{\pi}{k!} n_\mathrm{F}^{(k)}(\varepsilon_\alpha)
\ee
where $n^{(k)}_\mathrm F$ is the $k^\textrm{th}$ derivative of the Fermi function $n_{\textrm{F}}(\ep)=1/[e^{\beta(\ep - \mu)}+1]$.
In order to shorten the expressions, the susceptibility is written in units of $\frac{\mu_0 e^2 }{ \hbar^2}$ and the main text shorthand notations are introduced
\be 
n_\alpha \equiv n_F( \varepsilon_{\alpha}(\k)) \ , \   \langle \cdots \rangle_\mathrm{B Z} \equiv
\sum_{\alpha=\pm}   \int \cdots   \frac{d^2 k}{ 4 \pi^2} \ .
\ee
The different terms appearing in Eq. (\ref{UVb}) then give rise to the different susceptibility contributions:
\be
\chi_{\textrm{orb}}=\chi_{\textrm{LP}}+\chi_{\Omega}+\chi_g+\tilde \chi_g,
\ee 
with
\be
\chi_{\textrm{LP}}= \left\langle  \frac{n'_\alpha}{ 12}
 ( \partial_{xx}\varepsilon_{\alpha} \partial_{yy}   \varepsilon_{\alpha}  -   \partial_{xy}\varepsilon_{\alpha}\partial_{xy}
    \varepsilon_{\alpha}) \right\rangle_{\textrm{BZ}},
    \ee
\be
\chi_{\Omega}=\left\langle \left(-n_\alpha'+ \alpha \frac{n_\alpha }{ \ep}\right)  {\cal M}^2 \right\rangle_{\textrm{BZ}}, \qquad  {\cal M}= \ep \Omega \\
\label{supp-chiomega}
\ee
and
\be
\chi_g+\tilde \chi_g=
\left\langle \left( n_\alpha'- \alpha \frac{n_\alpha}{\ep}\right)   g^{ij} \partial_{i}\ep_{\alpha} \partial_{j} \ep_{-\alpha} -\alpha \frac{n_\alpha}{\ep}\frac{1}{2}
\partial_{ij}(\varepsilon^2 g^{ij}).
\right\rangle_{\textrm{BZ}}.
\label{supp-chig}
\ee
There are different ways to write the separate contributions $\chi_g$ and $\tilde \chi_g$.
The following compact expressions are obtained by performing integrations by part in order to eliminate all Fermi surface contributions:
\be
\chi_g=\left\langle - \alpha \frac{n_\alpha}{ \ep}
 Z_g \right\rangle_{\textrm{BZ}}, \qquad 
\tilde \chi_g=\left\langle - \alpha \frac{n_\alpha }{\ep}
\tilde  Z_{g} \right\rangle_{\textrm{BZ}}
\label{compact1}
\ee
with
\be
\label{compact2}
 Z_g=\frac{1}{2} \partial_{j}(\varepsilon^2\partial_{i}g^{ij})
, \qquad \tilde  Z_{g}=g^{ij}\partial_{i}\varepsilon_0 \partial_{j}\varepsilon_0+\alpha
 \varepsilon \partial_{i}(\partial_{j}\varepsilon_0 g^{ij}).
\ee
To conclude, it is worth mentionning that the distinct susceptibility contributions verify separately the sum rule over the full zero-field spectrum:
\be
\int \rm{d}\mu \chiLP(\mu)=\int \rm{d}\mu \chiOm(\mu)=\int \rm{d}\mu \chig(\mu)=\int \rm{d}\mu \chigtilde(\mu)=0.
\ee
Furthermore each susceptibility contribution vanishes for $\mu$ outside the full spectrum. 
These last two properties, together with the rather natural interpretation of each contribution (see below) strengthens 
the above decomposition of $\chiinter$ into three terms in comparison with the various decompositions adopted in other works (see below). \cite{Blount,Gao15,ogata2015} 

\medskip

\subsection{Role of symmetries on interband susceptibility contributions $\chiOm,\chig,\chigtilde$}

As already mentionned in the main text, for systems that are time-reversal invariant, 
inversion and particle-hole symmetries permit to discriminate the three interband susceptibility contributions $\chiOm,\chig,\chigtilde$.
For systems with an inversion symmetry, the Berry curvature vanishes ($\Omega(\k)=0$) and therefore $\chiOm(\mu)=0$ (From that perspective,
the case of graphene should be understood as the zero gap limit of a system that breaks inversion symmetry such as boron nitride). Similarily,
systems with particle-hole symmetry ($\varepsilon_0(\k)=0$) verify $\chigtilde(\mu)=0$.
The contribution $\chig(\mu)$ is thus the only one that remains when both inversion and particle-hole symmetries are simultaneously present.
In that respect it may be seen as the most fundamental one. 

Beyond global symmetries, it nevertheless appears that $\chi_g$ vanishes for systems such
that ${\bm h}(\k)$ only depends on either $k_x$ or $k_y$. For such a case, 
the metric tensor has a single non-vanishing component $g_{xx}$ or $g_{yy}$ and 
therefore $\chig$ and $\chiOm$ both vanish. In that situation the only non-vanishing component is $\chigtilde$. A simple example 
is a square lattice with nearest neighbor hopping $t$ and an alternating onsite potential $\pm \Delta$ along the $x$ direction. The
corresponding hamiltonian matrix is constructed from $\varepsilon_0=2t \cos(k_y)$ and ${\bm h}(\k)=(2t\cos(k_x),0,\Delta)$.


{\fred{
\section{Field induced positional shift in two-band systems}

Very recently it was shown that the presence of a magnetic field induces a band dependent positional shift 
of the Berry connection such that ${\bm a}_\alpha\rightarrow{\bm a}_\alpha+ B {\bm a'}_{\alpha}$, to linear order in magnetic field. \cite{Gao14,Gao15} (here  $a'(\k)$ 
does not contain the magnetic field, so that the full Berry connection is $a(\k) + a'(\k)B$.)
According to the semiclassical wavepacket formalism, \cite{Gao14,Gao15} the positional shift is composed of two distinct contributions
so that ${\bm a'}_{\alpha}(\k)$ can be written
\be  
{\bm a'}_{\alpha}=
{\bm a}_g+\alpha \tilde {\bm a}_g,
\ee
where the contribution ${\bm a}_g$ originates from the horizontal mixing of zero field band eigenstates 
whereas the contribution $\tilde {\bm a}_g$ comes from further vertical mixing of band eigenstates. \cite{Gao14,Gao15}
For two-band systems in two-dimension and in a perpendicular magnetic field, the positional shift contributions ${\bm a}_g$ and  $\tilde {\bm a}_g$  take the form
\be
{\bm a}_g(\k)=\frac{{\bm B}}{B} \times \frac{1}{2}
\left(
\ba{c}
\partial_j g^{xj}\\
\partial_j g^{yj}\\
0
\ea
\right),
\qquad
\tilde {\bm a}_g(\k)= \frac{{\bm B}}{B} \times {1 \over \varepsilon}
\left(
\ba{c}
g^{xj}\partial_j \ep_{0}\\
g^{yj}\partial_j \ep_{0}\\
0
\ea
\right),
\ee
Defining the corresponding field induced shifts of the Berry curvature,
\be
\Omega_g(\k)=[\nabla_\k \times {\bm  a}_g ]_z=-\frac{1}{2} \partial_i \partial_j g^{ij}
\ee
and
\be
\tilde \Omega_g(\k)=[\nabla_\k \times \tilde {\bm  a}_g ]_z=-\partial_i (\frac{g^{ij}\partial_j \ep_{0}}{\varepsilon}),
\ee
it is clear that each of them is an even function in $\k$-space $\Omega_g(\k)=\Omega_g(-\k)$ and similarily for $\tilde \Omega_g(\k)$; which is to be expected since 
it depends on the magnetic field that explictly breaks time-reversal symmetry. 
Despite this, each Berry curvature shift verifies the sumrule $\int_{\rm{B Z}}  {d^2 k} \  \Omega_{g}(\k)=\int_{\rm{B Z}}  {d^2 k} \  \tilde \Omega_{g}(\k)=0$. 

}}

\end{widetext}




\begin{thebibliography}{99}


\bibitem{AM} N. W. Ashcroft \AND    N. D. Mermin, {\it  Solid State Physics}, Saunders College, Fort Worth, (1976).
\bibitem{Landau30}L.D. Landau, Z. Phys. \textbf{64}, 629 (1930). See also the {\it Collected papers of L. D. Landau} edited by D. ter Haar (Pergamon Press, New York, 1965).
\bibitem{Peierls33}R. Peierls, Z. Phys. \textbf{80}, 763 (1933).
\bibitem{Adams52}E.N. Adams, Phys. Rev. \textbf{89}, 633 (1953).
\bibitem{Hebborn59}J.E. Hebborn \AND E.H. Sondheimer, J. Phys. Chem. Solids \textbf{13}, 105 (1960).
\bibitem{Roth61}L.M. Roth, J. Phys. Chem. Solids \textbf{23}, 433 (1962).
\bibitem{Blount} E.I. Blount, Phys. Rev. B  \textbf{126},   1636 (1962).
\bibitem{Wannier64}G.H. Wannier, Phys. Rev. \textbf{136}, A803 (1964).
\bibitem{Misra69}P.K. Misra \AND L.M. Roth, Phys. Rev. \textbf{177}, 1089 (1969).
\bibitem{Fukuyama71}H. Fukuyama, Prog. Theor. Phys. \textbf{45}, 704 (1971).
\bibitem{Fukuyama70}H. Fukuyama, J. Phys. Soc. Jap. \textbf{28}, 570 (1970).
\bibitem{Xiao10}D. Xiao, M.C. Chang \AND Q. Niu, Rev. Mod. Phys. \textbf{82}, 1959 (2010).
\bibitem{Thonhauser11} T. Thonhauser, Int. J. Mod. Phys B \textbf{25}, 1429 (2011).
\bibitem{Thonhauser05} T. Thonhauser, D. Ceresoli, D. Vanderbilt \AND R. Resta, Phys. Rev. Lett. \textbf{95}, 137205 (2005).
\bibitem{DiXiao05} D. Xiao, J. Shi \AND Q. Niu, Phys. Rev. Lett. \textbf{95}, 137204 (2005).
\bibitem{Gao15} Y. Gao, S.A. Yang \AND Q. Niu, Phys. Rev. B \textbf{91}, 214405 (2015).
\bibitem{Berry89} M.V. Berry, {\it The quantum phase, five years after} in {\it Geometric Phases in Physics}, A. Shapere and  F. Wilczek eds.,   World Scientific, Singapore (1989).
\bibitem{Vallee} J. P. Provost \AND  G. Vallee, Comm. Math. Phys. {\bf 76}, 289 (1980).
\bibitem{Marzari-Vanderbilt} N. Marzari and D. Vanderbilt, Phys. Rev. B {\bf 56}, 12847 (1997) 
\bibitem{RestaEPJB11} R. Resta, Eur. Phys. J. B {\bf 79}, 121137 (2011).
\bibitem{Neupert13} T. Neupert, C. Chamon and C. Mudry,  Phys. Rev. B {\bf 87}, 245103 (2013).
\bibitem{PeanoTorma15} S. Peotta, P. T\"orm\"a, Nature Com. {\bf 6}, 8944 (2015).
\bibitem{Inamoglu15} A. Srivastava and A. Imamo\u{g}lu, Phys. Rev. Lett. {\bf 115} 166802 (2015).
\bibitem{Lim15} L-K. Lim, J-N. Fuchs and G. Montambaux,  Phys. Rev. A {\bf 92}, 063627 (2015).
\bibitem{RaouxPRB15}  A. Raoux, F. Pi\'echon, J.-N. Fuchs \AND G. Montambaux, Phys. Rev. B  \textbf{91},  085120 (2015).
\bibitem{Gomez-Santos11}G. Gomez-Santos \AND T. Stauber, Phys. Rev. Lett. \textbf{106}, 045504 (2011).
\bibitem{RaouxPRL14}A. Raoux, M. Morigi, J.N. Fuchs, F. Pi\'echon \AND G. Montambaux, Phys. Rev. Lett. \textbf{112}, 026402 (2014).
\bibitem{Stauber15}A. Guti\' errez-Rubio, T. Stauber, G. G\` omez-Santos, R. Asgari \AND F. Guinea, Phys. Rev. B  \textbf{93},  085133 (2016).
\bibitem{Vignale91}G. Vignale, Phys. Rev. Lett \textbf{67}, 358 (1991).
\bibitem{Koshino10}   M. Koshino \AND T. Ando, Phys. Rev. B \textbf{81}, 195431 (2010).
\bibitem{Tarruell:12} L. Tarruell, D. Greif, T. Uehlinger, G. Jotzu \AND T. Esslinger, Nature \textbf{483}, 302 (2012).
\bibitem{Lim:12} L.-K. Lim, J.-N. Fuchs \AND G. Montambaux, Phys. Rev. Lett. \textbf{108}, 175303 (2012).
\bibitem{McClure56}J.W. McClure, Phys. Rev. \textbf{104}, 606 (1956).
\bibitem{Fuchs10}J.N. Fuchs, F. Pi\'echon, M.O. Goerbig \AND G. Montambaux, Eur. Phys. J. B \textbf{77}, 351 (2010).
\bibitem{Mielke91} A. Mielke, J. Phys. A {\bf 24}, 3311 (1991).
\bibitem{Aoki96} H. Aoki, M. Ando \AND H. Matsumara, Phys. Rev. B \textbf{54}, R17296 (1996).
\bibitem{Tasaki92} H. Tasaki, Phys. Rev. Lett. {\bf 69}, 1608 (1992)
\bibitem{Gao14}Y. Gao, S.A. Yang \AND Q. Niu, Phys. Rev. Lett. \textbf{112}, 166601 (2014). Here  $a'(\k)$ 
does not contain the magnetic field, so that the full Berry connection is $a(\k) + a'(\k)B$.
\bibitem{Savoie12} B. Savoie, J. Math. Phys. {\bf 53}, 073302 (2012).
\bibitem{ogata2015} M. Ogata and H. Fukuyama, J. Phys. Soc. Jap. \textbf{84}, 124708  (2015).


\end{thebibliography}
\end{document}